**Weakest-link control of invasive species: Impacts of memory, bounded rationality and network structure in repeated cooperative games.**

Adam Kleczkowski[1,2], Andrew Bate[3], Michael Redenti[2] and Nick Hanley[4]

[1] Dept. of Mathematics and Statistics, University of Strathclyde, Glasgow, United Kingdom

[2] Faculty of Natural Sciences, University of Stirling, Stirling, United Kingdom

[3] Department of Environment, University of York, York, United Kingdom

[4] IBAHCM, University of Glasgow, United Kingdom

Corresponding Author:

Adam Kleczkowski

Dept. Mathematics and Statistics, University of Strathclyde, Glasgow G1 1XH, United Kingdom

Email address: a.kleczkowski@strath.ac.uk

Keywords: Bioeconomic modelling; cooperation games; repeated games; networks; agriculture; crop protection; disease control; pest control.

JEL classification: C69, C72, C91, C92

**Abstract**

The nature of dispersal of many invasive pests and pathogens in agricultural and forestry makes it necessary to consider how the actions of one manager affect neighbouring properties. In addition to the direct effects of a potential spread of a pest and the resulting economic loss, there are also indirect consequences that affect whole regions and that require coordinated actions to manage and/or to eradicate it (like movement restrictions). In this paper we address the emergence and stability of cooperation among agents who respond to a threat of an invasive pest or disease. The model, based on the weakest-link paradigm, uses repeated multi-participant coordination games where players' pay-offs depend on management decisions to prevent the invasion on their own land as well as of their neighbours on a network. We show that for the basic cooperation game agents select the risk-dominant strategy of a Stag hunt game over the pay-off dominant strategy of implementing control measures. However, cooperation can be achieved by the social planner offering a biosecurity payment. The critical level of this payment depends on the details of the decision-making process, with higher trust (based on reputation of other agents reflecting their past performance) allowing significant reduction in necessary payments and slowing down decay in cooperation when the payment is low. We also find that allowing for uncertainty in decision-making process can enhance cooperation for low levels of payments. Finally, we show the importance of industry structure to the emergence of cooperation, with increase in the average coordination number of network nodes leading to increase in the critical biosecurity payment.

# 1. Introduction

There are many situations where failure to achieve an outcome depends not only on one's actions, but also the actions of others (Sims et al., 2016). For example, our efforts to prevent the spread of a disease can be undermined if someone else does not take appropriate precautions like hand washing, vaccination or protection (Kleczkowski et al., 2015; Maharaj and Kleczkowski, 2012). One particular case on which we focus in this paper involves biosecurity in agriculture and forestry (Macpherson et al., 2017, 2016). A lapse in biosecurity can result in the introduction and spread of animal and plant pests and pathogens, and can bring a large variety of negative consequences. These include the reduction in yield or quality of produce, reduced animal welfare, as well as the need to resort to costly and often unpopular reactive measures like widespread application of pesticides or culling. In addition, if one farm gets infected, biosecurity for others gets harder and so other farms are likely to follow, triggering a chain reaction (Fraser, 2016). Wider 'indirect' impacts from an infection that affect more than just the farms that initially get infected include regulatory and market impacts, such as trade bans, movement restrictions, culls and other biosecurity measures (Haydon et al., 2004; Knight-Jones and Rushton, 2013; Moslonka-Lefebvre et al., 2016).

An example where indirect damages might be highly important is the case of a bacterial pathogen, *Xylella fastidiosa* (Forestry Commission, 2018). It is a multihost plant pathogen (with 350+ suspected plant host species), spread by xylem-feeding insects. It is endemic in many places in the world but some European countries, including the UK are so far free from it (Forestry Commission, 2018). *X. fastidiosa* has been spreading in southern European countries, devastating hosts like olive, almonds and oleander (Martelli et al., 2016). Given the wide array of host, many of whom are asymptomatic, likely impacts include non-market ecosystem damages as well as commercial production losses. In countries like the UK the current main concern is not the direct damages, as the most vulnerable host species are not economically important and the xylem-feeding insects are less active with the cooler climate, but indirect damages. In the UK, the same EU crop protection regulations apply as those in

Italy or Spain where the pest is present, with long-term movement restrictions in the radius of 10km around known outbreaks, lasting for as long as 5 years (Forestry Commission, 2018).

For those that trade plants (e.g. plant nurseries) these regulations essentially mean destroying all hosts and a ban or severe restrictions in movement of host plants for a long period. Importantly, the damage is not limited to the trader that through negligence introduced the pathogen into the area, but to their neighbours over a potentially large geographic area. In addition, the response might also include contact tracing (similar to the one employed in controlling Foot and Mouth outbreak in the UK in 2001 (Haydon et al., 2004)) and so affect the sites located outside the immediate neighbourhood of the outbreak focus, but connected to it through trade. In short, if one agent employs lax biosecurity, other agents suffer.

Issues around biosecurity can be considered as a weakest link public good, where efforts to produce a desired level of biosecurity are undermined by a single actor in the population, even if all other actors implement the prevention measures (Perrings et al., 2002). This feature, together with the assumptions that investment in biosecurity is costly, but only worthwhile if successful, can be captured in a game theoretical framework. In this paper we use the Stag hunt game paradigm to describe the strategic interaction among decision makers threatened with an outbreak of plant diseases. The key feature of the Stag hunt game (Skyrms, 2013; Van Huyck et al., 1990) is the emergence of two Nash equilibria, one of which is pay-off dominant (when all players invest in biosecurity) and the other risk dominant (no players invest in biosecurity). We extend the concept of the Stag hunt game to a multi-player situation (Rich et al., 2005a) in which the interactions are described by a network with a mixture of spatial (local) links, representing geographic proximity, and random (non-local) links, representing trade movements between players (for an application of network theory to plant health and particularly *Xylella*, see (Jeger et al., 2007; Strona et al., 2017)).

The Stag hunt game has extensively been studied both theoretically (Büyükboyacı, 2014; van Veelen and Nowak, 2012; Weidenholzer and Simon, 2010) and in experiments (Banerjee et al., 2014, 2012), in a two-player and multi-player, network settings (Skyrms and Pemantle, 2009; van Veelen and Nowak, 2012), and as a single or repeated game. The key feature emerging from these studies is that cooperation emerges only under very specific conditions. In this paper we apply the repeated Stag hunt game model to address the emergence of cooperation among plant nursery managers faced with a business-threatening pest or disease (Rich et al., 2005a, 2005b). We particularly seek answers to the following key questions: *Given that cooperation is unlikely to emerge under the Stag hunt conditions, how much payment for implementing biosecurity is needed to achieve effective control of the disease in the landscape? Can the payment be lowered if we provide mechanisms for building up trust? How does the critical payment depend on uncertainty in pay-offs and boundedly- rational behaviour? How does behaviour depend on the industry structure, particularly the number of trading partners?*

## 2. Methods.

We describe here a set of decision-making agents who broadly represent plant nursery managers, but who could also represent farmers growing crops or animals, or forest managers. Nurseries are connected to other nurseries either through local (geographical) links or through trade which can span the whole population; both types of links can facilitate spread of an invasive pest or disease.

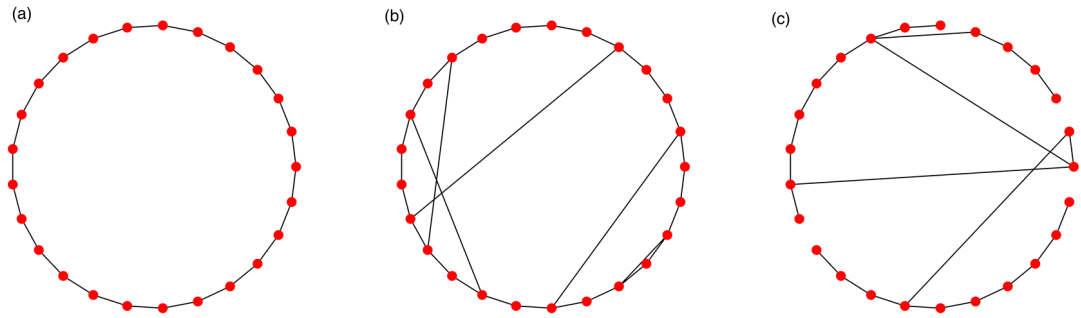

Fig. 1 Network structure: (a) regular ring network representing geographical connections, (b) the small-world network characterised by the addition of *m* long-range random links, and (c) the rewired network. For clarity, *M*=25 and *m*=5.

*The network.* We consider three network types. For the baseline case, we assume that the geographical structure is represented by a one-dimensional network with *M*=256 agents, *d*=2 nearest neighbours and periodic boundary conditions, i.e. a ring, see Fig. 1(a). This could represent, for example, a set of nurseries based along a major road or railway; the period boundary condition simply allows us to ignore the influence of the end-points (van Veelen and Nowak, 2012). We subsequently extend the model to represent non-local trade described by an addition of *m* random links connecting any node in the network to another node, Fig. 1(b). The resulting network has a small-world property (Jeger et al., 2007; Watts and Strogatz, 1998) with an increase in the average degree but large decrease in the average path length. For comparison, we also include a different small-world-type network, implementing the Watts-Strogatz rewiring algorithm (Watts and Strogatz, 1998); the network is formed from the basic ring structure by rewiring of (on average) *m* randomly selected links to anywhere in the population, Fig. 1(c). In such network the average degree does not change with *m* increasing, but the average path length decreases. The network is assumed to be static, with no response to the threats or changes in trading patterns (Gross et al., 2006).

*Basic game.* At each period of time (which could represent a year or a trading round), the agent is faced with a risk created by an invasive potential pest or disease which can be transmitted to the nursery from another place. Each agent can choose either to control/prevent the disease (C) or to do nothing (N). In our approach we do not model the actual spread of the pest but assume that the agent expects that if no control measure is taken on its premises or if at least one of the geographical or trading partners does not control the pest, then the plants at the nursery will be affected, resulting in the decrease in pay-offs. The consequences of the action C or N, both for the agent and for the neighbours, are realised in the same round and have no direct long-lasting consequences (except loss or gain of trust, see below).

*Pay-offs.* In each trading period, agents sell plants, with the healthy material bringing in the profit of *a* (arbitrary units) and the affected material bringing in *b* (arbitrary units), with $b < a$. The choice to control (C) is associated with a cost, *c*, independent of what neighbours do, while doing nothing (N) has no upfront cost. We additionally assume that the government tries to encourage the agents to control and in each period offers a subsidy *p* to each agent who chooses strategy C. The basic pay-off structure is given in Table 1.

Table 1. Basic pay-off structure. In simulations we choose $a = 16, b = 6, c = 6$ (arbitrary units) and the second value in this table corresponds to this choice.

| Neighbourhood:<br>Own action: | C (all neighbours control) | N (at least one neighbour does not control) |
|---|---|---|
| C (control) | $a+p-c=10+p$ | $b+p-c=p$ |
| N (do not control) | $b=6$ | $b=6$ |

*Repeated game.* We assume that the game is played repeatedly and the agents have a full knowledge of the decisions of their neighbours in the past. Each agent's actions are based on the expected pay-off which incorporates an estimate of what neighbours are going to do in the next step which we express in terms of two probabilities. $P_i(N=0)_{n+1}$ is the probability that in step *n*+1 all neighbours connected to the agent *i* will be implementing the control measures, whereas $P_i(N \geq 1)_{n+1} = 1 - P_i(N=0)_{n+1}$ represents the probability that at least one neighbour will not control. These probabilities can be evaluated assuming independence of neighbour decisions,

$$P_i(N=0)_{n+1} = \prod_{j \in nb_i} P_j(C)_{n+1}$$
$$P_i(N \geq 1)_{n+1} = 1 - \prod_{j \in nb_i} P_j(C)_{n+1}$$
(1)

where $nb_i$ is the set of neighbours of agent *i* and $P_j(C)_{n+1}$ represents the probability that the neighbour *j* adopts the control strategy in step *n*, as estimated by agent *i*.

*Estimation of $P_j(C)_{n+1}$.* In the model we assume that this is based on the information on the past performance of the agents. In the simplest case, this can be simply dependent on the last period; if any of the neighbouring agents selects strategy C, we can assume that it is likely to do this again in the next step. Let $x_n = 1$ if individual *j* played C at time *n*, and $x_n = 0$ otherwise. Then, $P_j(C)_{n+1} = x_n$ can be treated as an estimate that neighbour *j* will again choose the control strategy in the next step.

A natural extension of this process incorporates a form of trust-building (Bloembergen et al., 2015; Enright and Kao, 2015; Golman and Page, 2010; Skyrms, 2008) and bases the estimation of $P_j(C)_{n+1}$ on a past history of the agent *j*'s actions (Horváth et al., 2012). In our approach this is based on as a weighted memory of whether that neighbour implemented control previously, but with recent actions having a greater weighting, so that the estimated probability that individual *j* implements control in step *n*+1 is given by

$$P_j(C)_{n+1} = \frac{\sum_{k=0}^{n} x_k \, e^{-(n-k)/\tau}}{\sum_{k=0}^{n} e^{-(n-k)/\tau}} \quad (2)$$

where $x_k = 1$ if individual $j$ played C at time $k$, and $x_k = 0$ otherwise. Here, $\tau$ characterises how long past acts are remembered, with small values of $\tau$ resulting in high weight placed on recent turns, whereas a large value of $\tau$ means that acts take a long time to be forgotten. In this way, $\tau$ can be interpreted as a characteristic time over which the memory of past actions is kept. We arbitrarily assume that $\tau = 0$ corresponds to only the most recent event remembered, i.e. $P_j(C)_{n+1} = x_n$.

*Calculation of the pay-offs.* Once the probabilities in Eq. (2) are calculated, the expected pay-offs of the agent $i$ can be computed as follows

$$\begin{aligned} E_i[C]_{n+1} &= P_i(N=0)_{n+1} \times (a+p-c) + P_i(N \geq 1)_{n+1} \times (b+p-c) \\ E_i[N]_{n+1} &= P_i(N=0)_{n+1} \times b + P_i(N \geq 1)_{n+1} \times b \end{aligned} \quad (3)$$

where $E_i[C]_{n+1}$ is the expected pay-off if the agent $i$ chooses strategy C and $E_i[N]_{n+1}$ if chooses strategy N (note that the second expression simplifies to $b$).

*Pure and mixed strategies.* Deterministically, the agent will choose the action that has the highest expected pay-off, i.e. will always choose strategy C if $E_i[C]_{n+1} \geq E_i[N]_{n+1}$ and strategy N if $E_i[C]_{n+1} < E_i[N]_{n+1}$. However, this assumes that the agent possesses complete knowledge of the pay-offs and outcomes. In order to represent the uncertainty we use the idea of quantal response learning (Harsanyi, 1973; Mckelvey and Palfrey, 1998; Rosenthal, 1989). At each time step the agent $i$ evaluates the expected pay-offs as described above but then chooses to play C randomly with probability

$$\Pr(C)_{n+1} = \frac{1}{1+\exp(-\lambda(E[C]_{n+1} - E[N]_{n+1}))} \quad (4)$$

and N otherwise. This function varies with $\lambda$ between two extremes, the deterministic choice based on highest pay-off when $\lambda \to \infty$, and the indifference to expected pay-offs (i.e. a decision based on a coin-flip) for $\lambda = 0$ when $\Pr(C)_{n+1} = 1/2$; the chance of choosing to play C (N) increases as the expected pay-off for playing C (N) increases, see Fig. 2.

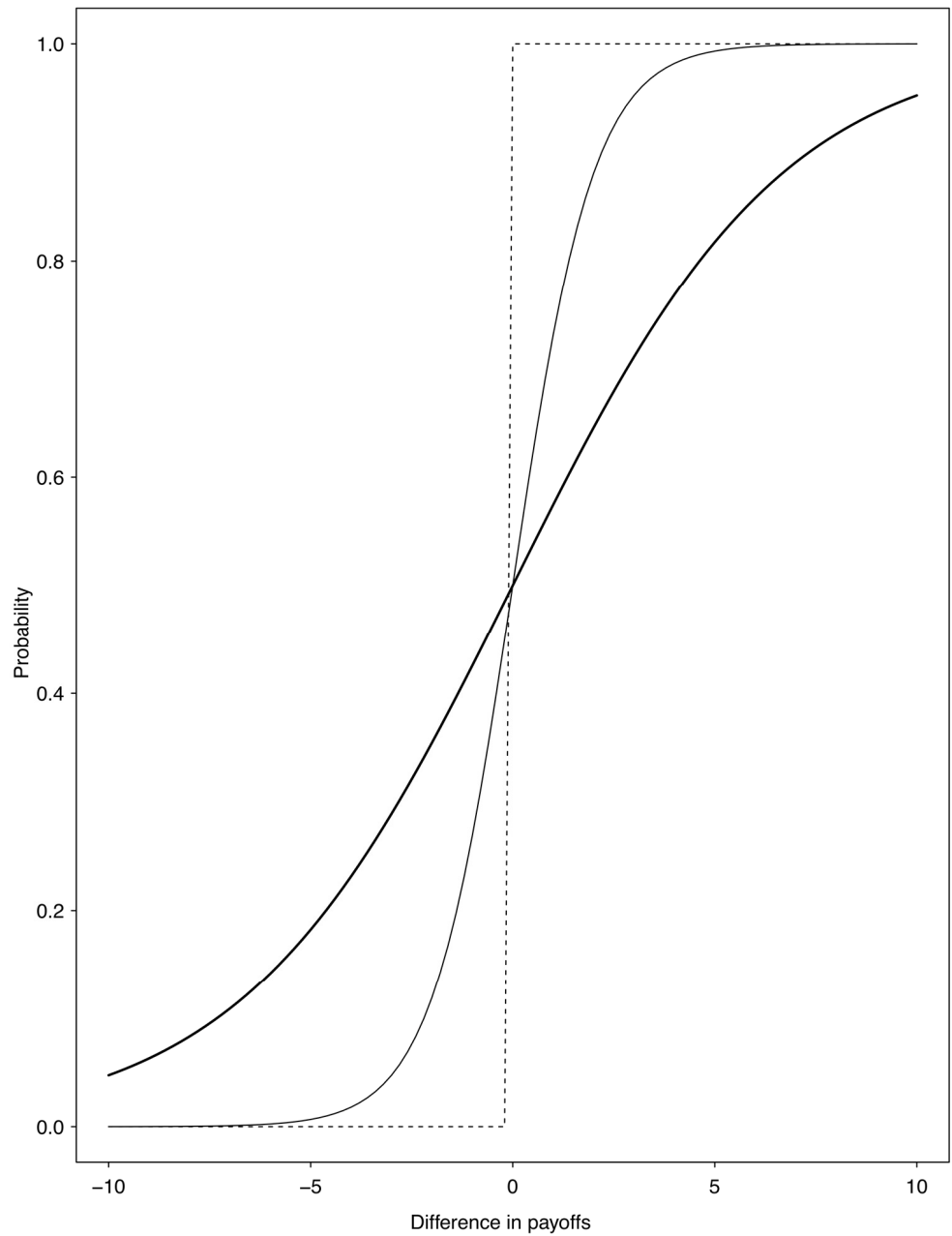

Fig. 2: Graph of the function given by Eq. (4), the probability of choosing the control strategy, as a function of difference in pay-offs. The dashed line illustrates the deterministic case of $\lambda \to \infty$, the thin solid line represents $\lambda = 1$, and the thick solid line to $\lambda = 0.2$.

*Initial conditions.* We assume that at time $n=0$ a new (invasive) pest poses a threat to the agents and then study how cooperation evolves in response to this threat. We consider two different initial conditions, one where most (80%) agents cooperate at the start of the model, and one where most (80%) agents initially do not implement control (i.e. 20% implement control). The first case corresponds to the situation when the cooperation is a default behaviour and we study whether it survives the change in the conditions. In the second case, we study whether the full cooperation will emerge over the long time.

*Parameters.* The pay-off values in the paper are arbitrary and for simulation purposes we choose $a=16, b=6, c=6$ so that the profit from the infected material exactly balances the cost of control and the profit from the healthy material is higher than the profit from the infected material (note that the model considered here is invariant under the addition of the same constant to *all* terms in Table 1 simultaneously). Typically, 128 simulations are carried out for the network size of $M=256$ agents, with $T=1024$ repeated games starting with random initial conditions and involving random decisions based on Eq. (4), if applicable.

## 3. Results

A successful prevention or control of invasive diseases requires coordination of agent actions across the landscape. In this paper, we take it as our objective that as many agents as possible choose the control option (C) over the 'do nothing' option (N), given the level of payment, $p$. We particularly address the dependence of the critical biosecurity payment needed to achieve cooperation, $p_c$, on agents' memory $\tau$, the parameter capturing the bounded rational behaviour of the agents, $\lambda$, and the number of long-range links, $m$.

*Basic model.* In absence of any additional payment ($p=0$) the two-agent, single-period model corresponds to a Stag hunt game as long as $a-b+p>c, p<c$; that is when the difference in profits when selling healthy and diseased stock plus the payment for biosecurity are strictly larger than the cost of control. Which strategy is chosen depends on how the agent estimates the probability that the other player selects C (or N), $P_j(C)_{n+1}$. In the simplest case of $P_j(C)_{n+1} = 1/2$, strategy N is risk dominant for $p < 1$. However, in the 'worst-case' scenario, the agent needs to expect other players to fail to control the pest. Thus, $P_j(C)_{n+1} = 0$ and strategy N is risk dominant for $p < c$.

For the network model, if the decision is based on the last choice made by the neighbours, $\tau = 0$, and under the deterministic decision process, $\lambda \to \infty$, the result is similar to the 'worst-case' scenario of the simple two-agent game, with the critical biosecurity payment of $p_c = c$. If $p < c$ then the control strategy ceases to be adopted as time passes, independently of the initial condition, Fig. 3(a). This results from a contagious nature of strategy N: The knowledge that one of the agent neighbours selected N in the last step and therefore is expected to do it again makes the expected pay-off for cooperating worse than not knowing and just guessing what the neighbour(s) will do randomly. Thus, once the agent believes others are going to play N, it plays N in response. For $p \geq c$, the pay-off for selecting C is larger than selecting N even if $P_j(C)_{n+1} = 0$ and so the number of agents adopting the control strategy increases until all agents cooperate (not shown here, but similar to Fig. 3(d)).

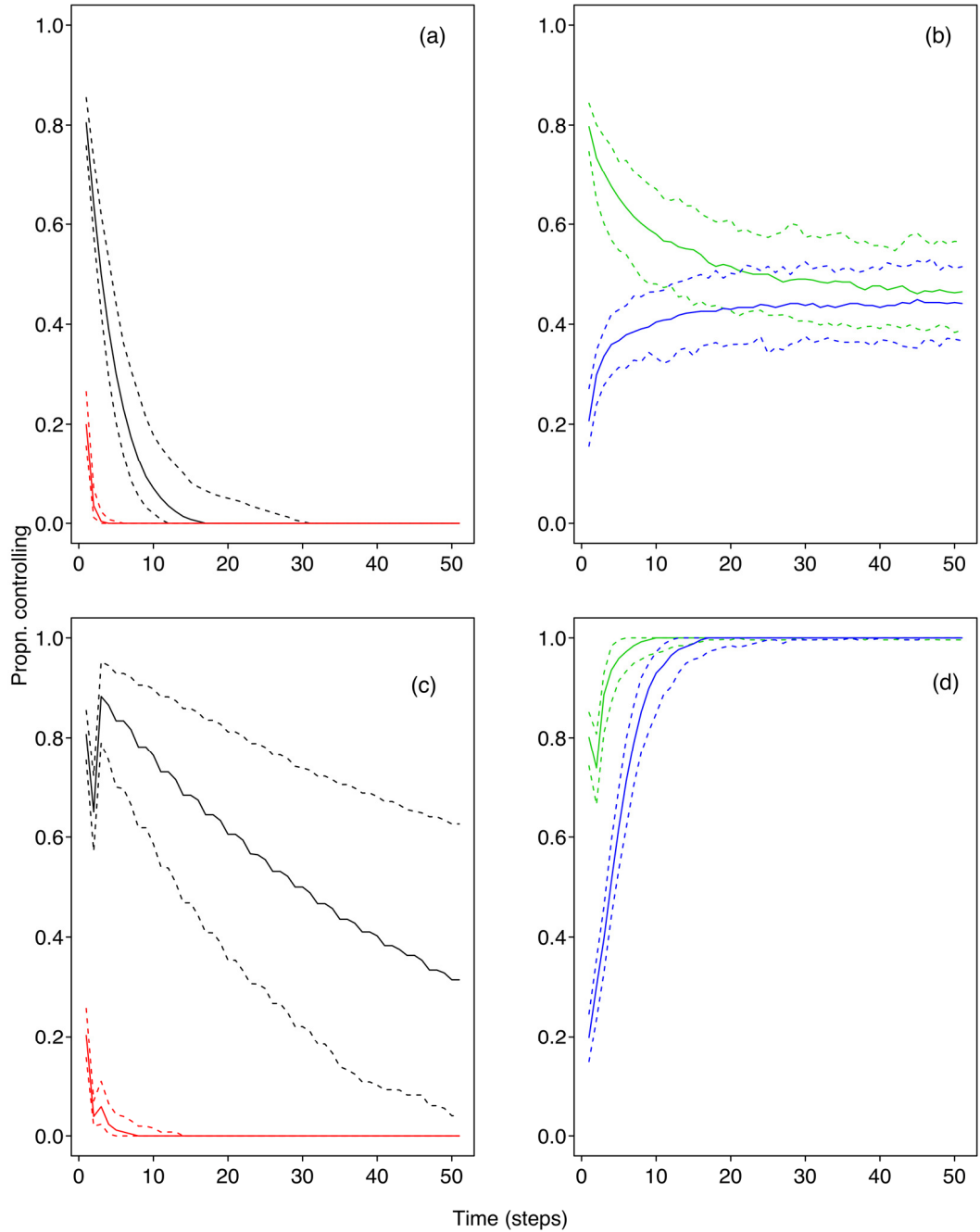

Fig. 3. The proportion of agents adopting the control strategy as a function of the number of repeated games, for $\lambda \to \infty$ (a) and (c) and for $\lambda = 1.0$ (b) and (d); $\tau = 0$ in (a) and (b), $\tau = 1$ in (c) and (d). System size is 256, with no long-range links. Solid lines represent median and dashed lines 2.5% and 97.5% quantiles simulated over 256 runs. Top set of curves corresponds to the initial number of agents adopting C of 80%, bottom set of curves for 20%. Biosecurity payment $p=5$ is below the critical value for the simple game, $c = 6$.

*Bounded rationality.* As $\lambda$ decreases, the agents stop always adopting the strategy with the higher expected pay-off and start occasionally making 'mistakes'. For moderate $\lambda = 1$ (while the decision is still based only on the last state, $\tau = 0$), the effect is twofold. Firstly, boundedly-rational behaviour leads to an increased range of values for the biosecurity payment *p* where cooperation levels in adopting control are high, Figs. 3(b) and 4(a). Secondly, some levels of cooperation persist for even smaller values of *p*, Fig. 3(b) and Fig. 4(a), although disappear as *p* declines to 0, Fig. 4(a). This counterintuitive behaviour emerges because the 'mistakes' allow clusters of cooperation to appear and to persist. The key to the stability of a cluster formed of agents adopting the C strategy is the perception of risk of the disease spreading amongst individuals at the end points of the cluster. They are likely to switch to N if they expect their neighbours to adopt N in the next step. However, if *p* is close to *c*, the difference between pay-offs $E_i[C]_{n+1}$ and $E_i[N]_{n+1}$ is small and so even if it is strictly speaking more profitable to adopt N, the agents are likely to still choose C (see Fig. 2).

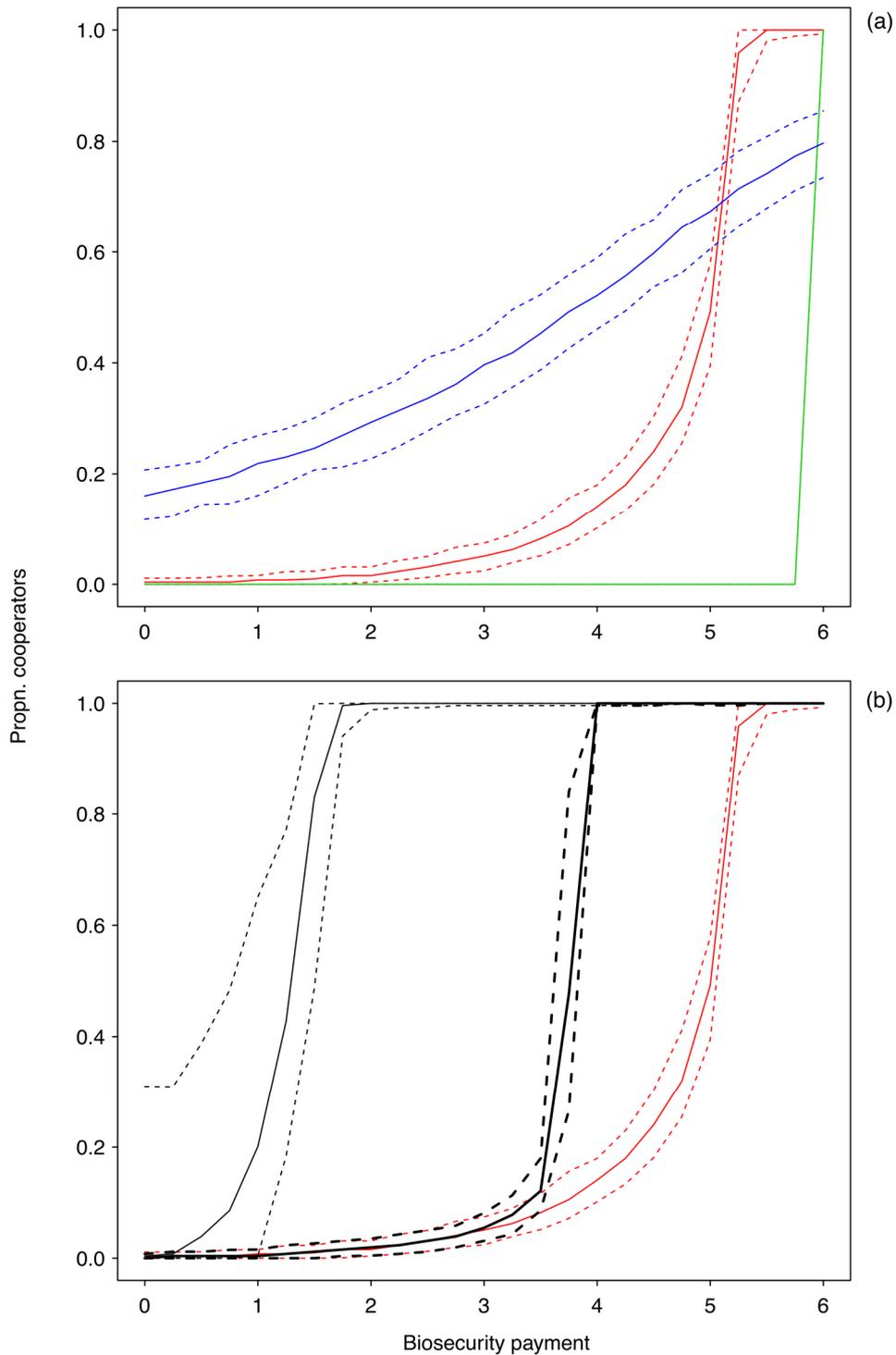

Fig. 4: The proportion of agents adopting the control strategy as a function of biosecurity payment, $p$, for (a) no memory ($\tau = 0$) but different values of $\lambda \to \infty$ (green), $\lambda = 1$ (red) and $\lambda = 0.2$ (blue), and (b) for $\lambda = 1.0$ with different memory levels, $\tau = 0$ (red), $\tau = 1$ (thick black) and $\tau = 10$ (thin black); the red curves are the same in (a) and (b). M=256, with no long-range links. Solid lines represent the median and dashed lines 2.5% and 97.5% quantiles simulated over 128 runs after 2048 periods. Initially 80% of the agents adopt control; the result for 20% is very similar after the transient behaviour shown in Fig. 3.

If $\lambda$ is reduced even more, the agents will display increasingly random behaviour without paying attention to what the neighbours have been doing. We find that full cooperation is no longer possible, even if the biosecurity payment is increased to *p=c*, but likewise some cooperation occurs even in the absence of the biosecurity payment, *p*, Fig. 4(a). As $\lambda \to 0$, the proportion of agents selecting strategy C tends to 1/2 independently of other parameters or the biosecurity payment, *p*.

*Memory and trust.* As the agents start paying more attention to the past events so that $\tau > 0$, the initial condition plays a more important role and the variability in the temporal path increases, Fig. 3(c), with some configuration of clusters of cooperation taking a very long time to convert to the N strategy.

When memory is short (small $\tau$), the main effect is to shift the critical value of the biosecurity payment, $p_c$, towards low values, thus enhancing cooperation, cf. Figs. 3(c) and 3(d); summarised in Fig. 4(b). Note that this requires some level of 'bounded rationality' to allow the neighbours of those who played N for a long time to consider playing C, as $p_c = c$ for any value of $\tau$ if $\lambda \to \infty$. For high levels of memory retention, $\tau = 10$, the region of biosecurity payment leading to consistent high cooperation levels is considerably widened, Fig. 4(b). Thus, even if the agent neighbour plays N once after having played C for a long time, its neighbours are likely to ignore this and continue playing C; this leads to a reduced 'contagion' of the N strategy, but also requires some level of bounded-rational behaviour.

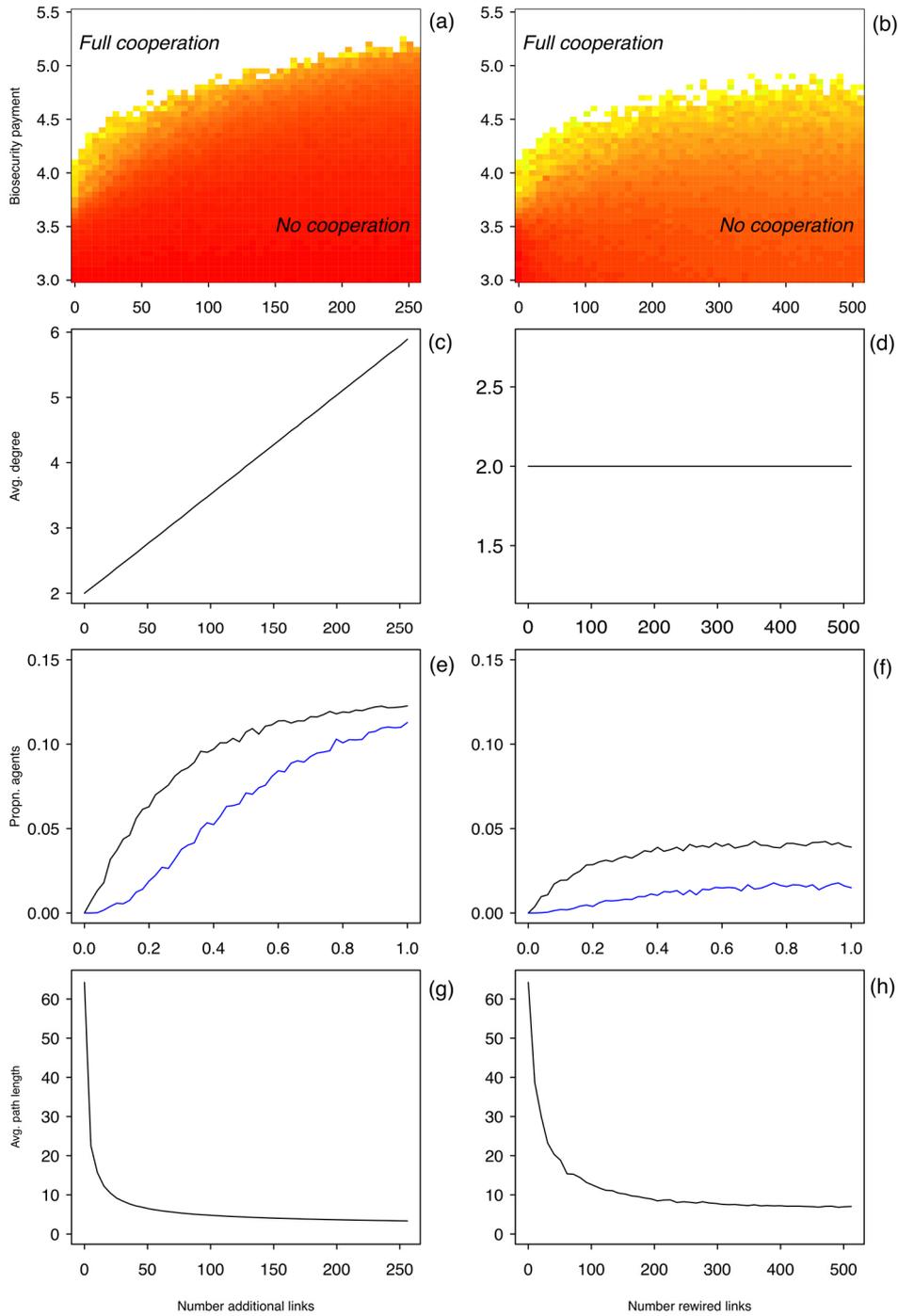

Fig. 5: (a) and (b): The average proportion of agents choosing to control after $T=1024$ steps as a function of the number of random long-range links and the biosecurity payment; red colour represents no cooperation, orange to yellow represents increasing cooperation, and white corresponds to complete cooperation. (c) and (d): the average degree, (e) and (f) the proportion of agents with 3 or more connections (black) and 4 or more connections (blue), and (g) and (h) the average path length for networks used in (a) and (b). Left column corresponds to small-world networks formed by addition of long-range links and right column to rewired small-world networks. Other parameters: $M=256$, $\lambda=1$ and $\tau=1$; the initial proportion of cooperators is 80% and the number of replicates is 25.

*Trade and random links.* The results so far assume a network of interactions with every agent influenced by their two nearest neighbours. As the number of additional long-range links representing trade increase, the critical biosecurity payment needed to encourage cooperation in disease control also increases for the addition network, Fig. 5(a). This is to be expected, as the additional links increase the average degree, Fig. 5(c), which means more agents are needed to adopt the control strategy for the population-level cooperation to succeed. However, the rewired network has a constant average degree, Fig. 5(d) with 5(c), but also shows the loss of cooperation as the proportion of non-local links increases, Fig. 5(b). Both networks are characterised by an increasing proportion of agents with 3 or more (and 4 or more) connections, Figs. 5(e) and 5(f). Thus, the loss of cooperation appears to be primarily associated with the increase in the proportion of agents who have more links. Such agents are less likely to select strategy C, as they perceive a higher risk of damage from trading links.

**4. Discussion and Conclusions.**

High risks of invasive pests and diseases like *X. fastidiosa* often lead authorities to impose draconian consequences if the disease agent is found on the premises. Moreover, not only those directly involved are affected by the emergency control measures, but also their neighbours will likely lose profits (and in the extreme, livelihoods). In this paper we have developed a modelling framework based on the concept of a weakest link public good and used it to examine conditions under which cooperation can emerge in the population of agents potentially affected by the spread of a pest or disease.

The lack of control in our model is a very strong transferable externality, as even a single person who does not implement biosecurity measures can cause huge losses across the whole industry (Shogren and Crocker, 1991). Hence, a rational response of agents to the disease threat is to mistrust their neighbours and assume the worst outcome – if any of them decides not to control the pest, all will be affected. As a result, agents tend to give up the control strategy, unless the state subsidy for biosecurity, *p*, fully compensates them for the costs of control, *c*. Hence, even if we start with a population in which almost all agents initially control, the agents most in contact with those that do *not* control will find that the best strategy is to cease to implement precautionary control, eventually resulting in the loss of cooperation in the whole population.

In the paper, we have explored ways in which behavioural mechanisms (bounded rationality, trust) can lower the critical threshold and, finally, what the role of long-range trade is. We found that both trust and bounded rationality tend to lead to higher levels of cooperation, even if the biosecurity payment is lower than the costs of control. In particular, the dependence of the agent's choice of strategy on the past decisions of their neighbours significantly increases the chances that the cooperation will emerge in the whole system, with the critical values of the biosecurity payment decreasing with increasing the characteristic time for the memory, $\tau$. However, even if $p < p_c$ for a given value of $\tau$, cooperation can persist for a long time, as exemplified in Fig. 3(c). Interestingly, the persistence time (defined here as the time needed for the proportion of agents adopting the control strategy to drop from 80% to 20%) scales with $\tau$ according to a power law, $T_{0.2} = 5.3 \times \tau^{1.8}$, Fig. 6. Thus, a relatively modest increase in $\tau$ can lead to a disproportionate increase in the persistence of cooperation. This effect stresses the importance of mechanisms that help to build trust in a framework that is inherently built on mistrust (as the agents expect to be affected even if only one neighbour does not implement control).

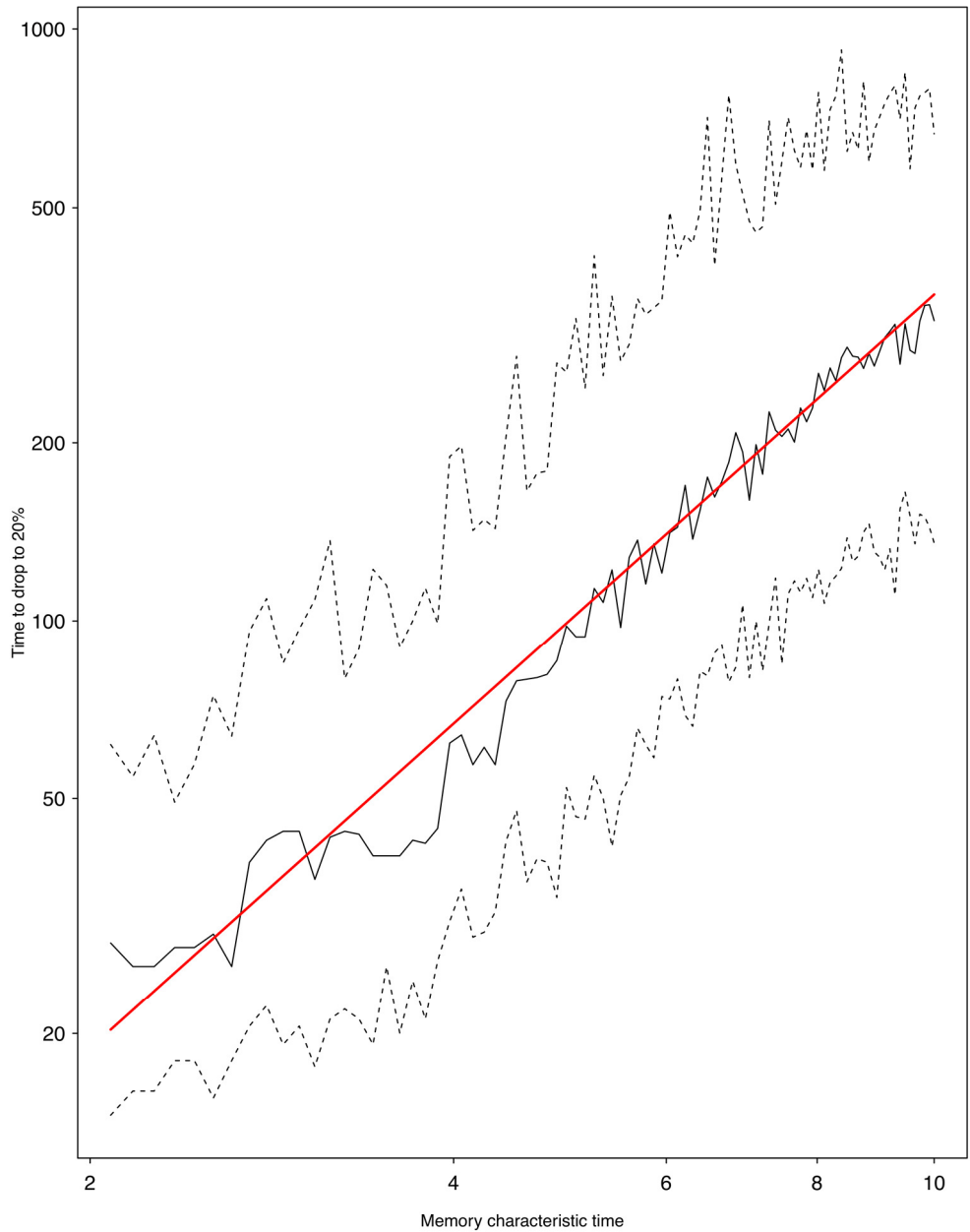

Fig. 6: The dependence of the time taken for the proportion of agents adopting the control strategy to drop from the initial 80% to 20%, $T_{0.2}$, on the characteristic time for memory, $\tau$, Eq. (2); solid line represents the median and the dashed lines the 95% confidence intervals based on 64 replicates. Note the double logarithmic scale, with a straight line in red representing the least square fit to all points, $\ln(T_{0.2}) = \alpha + \beta \ln(\tau)$ with $\alpha = 1.674 \pm 0.025$ and $\beta = 1.824 \pm 0.0144$, $F$=16,118 (5666df), $R^2 = 0.74$. Other parameters: $p$=0, $M$=256, $m$=0, $\lambda \to \infty$, $T$=1024.

Interestingly, the increased potential for 'errors' in calculating expected profits corresponding to decreasing $\lambda$ also makes cooperation more likely, unless the difference in pay-offs is high (Gächter, 2017). This can be associated with 'forgiveness', as the agents are more likely to forget one period when a neighbour 'cheated' and did not implement control, assuming the loss is not very big. This mechanism offsets the contrasting behaviour when the agent makes a mistake in the other direction and stops controlling even though all its neighbours choose to control. This also points in the direction of future research that would attempt to capture in more detail the decision-making process of agents faced with invasive disease or pest threats.

Industry structure also plays an important role in determining whether cooperation emerges or persists in the population, given our interpretation of industry structure as representing the number of trading links between farmers/foresters. Our results show that this is primarily driven agents who have more trading "neighbours" being less likely to adopt the control strategy. Those individuals then play a similar role to non-cooperators in Enright & Kao, (2015), although in our model there is no additional heterogeneity in their behaviour. More work is needed to identify whether the network topology, as exemplified here by the average path length, Fig. 5(g) and 5(h), also influences the dynamics of cooperation. In addition, in our model the network structure is fixed, whereas in reality, the agents might respond to disease threat by rewiring the trading pattern in the repeated game; such dynamic and adaptive networks have been a subject of intense research (Gross et al., 2006; Maharaj and Kleczkowski, 2012) and our model can be extended in this direction.

In our model, the process of actual disease spread is not explicitly incorporated in the framework, as we only look at the response of the agents to the disease threat (rather than the disease presence). This is probably the reason why network topology does not seem to play as important a role as the number of neighbours. An obvious extension of the model would be to incorporate an explicit epidemiological component to the model which represents the process of pathogen or pest spread and growth. Another extension would be incorporation of heterogeneity in the agent pay-offs (e.g. the profit from selling the healthy stock, $a$, or the cost of control, $c$) in addition to the number of connections considered in this paper.

**Acknowledgements**. This work was funded by the Tree Health and Plant Biosecurity Initiative (phase 2) grant by BBSRC, Defra, ESRC, Forestry Commission, NERC and Scottish Government, BB/L012561/1, the BBSRC project BB/M008894/1 and the NSF Grant 1414374 as part of the joint NSF-NIH-USDA Ecology and Evolution of Infectious Diseases program.